# Thermal imaging of spin Peltier effect

Shunsuke Daimon[1,2], Ryo Iguchi[1], Tomosato Hioki[1], Eiji Saitoh[1-5] and Ken-ichi Uchida[1,3,6]

[1]Institute for Materials Research, Tohoku University, Sendai 980-8577, Japan

[2]WPI Advanced Institute for Materials Research, Tohoku University, Sendai 980-8577, Japan

[3]Center for Spintronics Research Network, Tohoku University, Sendai 980-8577, Japan

[4]Spin Quantum Rectification Project, ERATO, Japan Science and Technology Agency, Sendai 980-8577, Japan

[5]Advanced Science Research Center, Japan Atomic Energy Agency, Tokai 319-1195, Japan

[6]PRESTO, Japan Science and Technology Agency, Saitama 332-0012, Japan

**When a charge current is applied to a junction comprising two different conductors, its temperature increases or decreases depending on the direction of the charge current. This phenomenon is called the Peltier effect[1,2], which is used in solid-state heat pumps and temperature controllers in electronics. Recently, in spintronics[3,4], a spin counterpart of the Peltier effect was observed[5]. The "spin Peltier effect" modulates the temperature of a magnetic junction depending on the direction of a spin current[6,7]. Here we report thermal imaging of the spin Peltier effect; using active thermography technique[8,9], we visualize the temperature modulation induced by a spin current injected into a magnetic insulator from an adjacent metal. The thermal images reveal characteristic distribution of spin-current-induced heat sources, resulting in the temperature change confined only in the vicinity of the metal/insulator interface. The finding of this anomalous temperature distribution allows us to estimate the actual magnitude of the temperature modulation induced by the spin Peltier effect, which is found to be more than one order of magnitude greater than that previously believed.**

In the field of spintronics, the interplay between spin, charge, and heat currents has been extensively studied from the viewpoints of fundamental physics and thermoelectric applications[10,11]. One of the triggers of this research trend is the discovery of the spin Seebeck effect (SSE)[12-15], which refers to the spin-current generation in linear response to a temperature gradient. The SSE is usually measured in a junction comprising ferromagnetic and paramagnetic materials; when a temperature gradient is applied across the junction, a spin current is generated due to a nonequilibrium state of spins at the ferromagnet/paramagnet interface[16-18], which in turn produces a measurable electric field in the paramagnet via the spin-orbit interaction[19,20]. The nonequilibrium state is excited by a thermally activated collective dynamics of magnetic moments, i.e. magnons[21], in the ferromagnet, a mechanism different from other spin-transport phenomena driven by conduction electrons[22,23]. Due to the magnon-driven nature, the SSE appears not only in conductors but also in insulators[13,15], expanding the scope of spintronics and thermoelectric technology[10,24].

The spin Peltier effect (SPE), the reciprocal of the SSE, refers to the heat-current generation in linear response to spin-current injection. The first observation of the SPE was reported by Flipse et al. in 2014 using a junction comprising a ferrimagnetic insulator yttrium iron garnet (YIG) and a paramagnetic metal Pt[5]; since the SPE appears in insulators, its physics is also discussed in terms of nonequilibrium magnon excitation in the same manner as the SSE[5,10,25]. Despite its scientific interest and technological importance, except for this demonstration, no experimental studies on the SPE have been reported so far due to difficulty in measuring this phenomenon. The difficulty mainly comes from the length scale of a spin current. Since a spin current disappears within a very short distance typically ranging from several nanometers to micrometers[26-28], the heat current due to the SPE is generated only in the range of the spin-current decay length. In ref. 5, to detect the temperature change in such a small scale, micro-fabricated thermopile sensors consisting of a number of thermocouples are integrated into a Pt/YIG-based device by using nanolithography. However, to further clarify the behavior of the SPE, a different approach is necessary.

In this study, we report the thermal imaging of the SPE in paramagnetic metal (PM)/ferrimagnetic insulator (FI) junction systems, which makes it possible to visualize the spatial distribution of spin-current-induced temperature modulation. This is realized by means of active infrared emission microscopy called lock-in thermography (LIT)[8,9]. Although a typical temperature change induced by the SPE is smaller than detection limits of conventional steady-state thermography, the LIT provides a much better signal-to-noise ratio and higher sensitivity with the temperature resolution of < 0.1 mK, enabling contact-free measurements of the spatial distribution of the SPE signals over a large area. In the LIT measurements, one inputs a periodic external perturbation, i.e. spin-current injection in our experiments, to a sample and extracts thermal images oscillating with the same frequency as the perturbation. The obtained thermal images are transformed into the lock-in amplitude and phase images by Fourier analysis. Here, the phase image gives the information about the sign of the temperature modulation depending on the periodic perturbation as well as the time delay due to thermal diffusion. Based on this technique, we reveal the distribution of heat sources induced by the SPE.

To excite the SPE, we employ the spin Hall effect (SHE)[20,26] as a tool to inject a spin current. Figure 1b shows a schematic illustration of the SHE-induced SPE in a PM/FI junction. When a charge current $\mathbf{J}_c$ is applied to PM with strong spin-orbit interaction, e.g. Pt or W, the SHE generates a spin current. This spin current then induces spin accumulation near the PM/FI interface, of which the spin-polarization vector $\boldsymbol{\sigma}$ is directed along $\mathbf{J}_c \times \mathbf{n}$ with $\mathbf{n}$ being the normal vector of the interface plane. When the spin accumulation combines with magnetic moments in FI via interfacial exchange interaction[29], it transfers spin-angular momentum and energy from electrons in PM to magnons in FI, or vice versa, across the interface as spin-transfer torque[6,29]. This spin/energy transfer process is proportional to the magnitude of the injected spin current and dependent on whether the $\boldsymbol{\sigma}$ direction in PM is parallel or antiparallel to the magnetization $\mathbf{M}$ of FI. The nonequilibrium state of the magnon and electron systems induces a heat current $\mathbf{J}_q$ across the PM/FI interface, which satisfies the following symmetry:

$$\mathbf{J}_q \propto (\boldsymbol{\sigma} \cdot \mathbf{M})\, \mathbf{n} \propto \mathbf{J}_c \times \mathbf{M} \tag{1}$$

Therefore, by measuring the spin-dependent temperature change in linear response to $\mathbf{J}_c$ near the PM/FI interface, the SPE can be demonstrated.

Figure 1a shows a schematic illustration of the sample system used in this study. The sample consists of a U-shaped Pt or W film formed on an YIG substrate (see Methods for details). To observe the SPE using the LIT technique, we measured the spatial distribution of infrared radiation thermally emitted from the sample surface while applying an a.c. charge current with rectangular wave modulation (with the amplitude $J_c$ and frequency $f$) to the Pt or W layer and extracted the first harmonic response of detected signals, where we set $f$ = 5 Hz except for $f$ dependent measurements (Fig. 1c and Supplementary Fig. 1). Here, the detected infrared radiation is converted into temperature information through the calibration method detailed in Supplementary Note 1. To enhance infrared emissivity and to ensure uniform emission properties, the sample surface was coated with insulating black ink, of which the emissivity is > 0.95. During the LIT measurements, an in-plane magnetic field **H** (with the magnitude $H$) was applied along the $x$ direction. When $|H| > 50$ Oe, **M** of YIG is aligned along the **H** direction. In our sample, owing to the U-shaped structure of the PM layer, the symmetry of the SHE can be confirmed simultaneously because the relative orientation of $\sigma$ and **M** is different between the areas L, R, and C, where $\sigma \parallel$ **M** on L and R and $\sigma \perp$ **M** on C (Fig. 1a,b). Because of the symmetry of the SHE-induced spin-transfer torque[26,29,30], the SPE signal can appear on L and R, while it should disappear on C (see equation (1)). Since the **J**$_c$ and resultant $\sigma$ directions on L are opposite to those on R, the sign of the SPE-induced temperature modulation is reversed between these areas. Importantly, by extracting the first harmonic response of the detected thermal images, we can separate the SPE contribution ($\propto J_c$) from Joule-heating contribution ($\propto J_c^2$), because the Joule heating generated by the rectangular a.c. current is constant in time as depicted in Fig. 1c. All the measurements were carried out at room temperature and atmospheric pressure.

In Fig. 2a,b, we show the LIT amplitude $A$ and phase $\phi$ images for the Pt/YIG sample at $J_c = 4.0$ mA, where the amplitude and phase are defined in the range of $A > 0$ and $0° \le \phi < 360°$, respectively. As shown in the $A$ image for $H = +200$ Oe (the left image of Fig. 2a), the clear temperature modulation depending on the **J**$_c$ direction appears on L and R, while it disappears on C, which is consistent with the aforementioned symmetry of the SHE (see equation (1)). Significantly, the difference in lock-in phases between L and R was observed to be ~180° (the left image of Fig. 2b); the input charge current and output temperature modulation oscillate with the same (opposite) phase on L (R) in the Pt/YIG sample at $H = +200$ Oe. Since the heat-conduction condition is the same between L and R, this $\phi$ shift is irrelevant to the time delay caused by thermal diffusion, indicating that the sign of the temperature modulation on the Pt/YIG surface is reversed by reversing the **J**$_c$ direction. We also found that the magnitude of $A$ is proportional to $J_c$, while the $\phi$ shift of ~180° remains unchanged with respect to $J_c$ (Fig. 2c-e); the temperature modulation appears in linear response to the charge current applied to the Pt layer.

Next, we measured the $H$ dependence of the LIT thermal images using the same Pt/YIG sample. When the **H** direction is reversed and its magnitude is greater than the saturation field, no $A$ change and clear $\phi$ reversal appear on L and R (the right images of Fig. 2a,b). At around $H = 0$ Oe, the thermal images exhibit patchy patterns corresponding to the magnetic domain structure of YIG, and the net temperature modulation averaged over the Pt/YIG surface disappears (the center images of Fig. 2a,b). These behaviors result in the odd $H$ dependence of the temperature modulation reflecting the magnetization process of YIG (Fig. 2f,g), consistent with the feature of the SPE.

To further support our interpretation that the current-induced temperature modulation originates from the SPE, we performed control experiments. In Fig. 3a,b, we show the LIT images for the W/YIG sample, where the $\sigma$ direction of the spin current flowing across the W/YIG interface is opposite to that across the Pt/YIG interface since the sign of the spin Hall angle of W is opposite to that of Pt[26,30]. On L and R of the W/YIG sample, the clear temperature modulation was observed. Significantly, the signal in the W/YIG sample is opposite in sign to that in the Pt/YIG sample (compare Figs. 2a,b and 3a,b), confirming that the temperature modulation originates from the SHE and spin-current injection across the PM/YIG interface. We also checked that the signal disappears in a Pt/Al$_2$O$_3$/YIG sample, where the Pt and YIG layers are separated by a thin film of insulating Al$_2$O$_3$ (Fig. 3c,d), indicating an essential role of the direct PM/YIG contact. These experiments, summarized in Fig. 3e,f, clearly show that the temperature modulation near the PM/YIG interfaces is attributed to the SPE driven by the SHE. We note that the sign of the temperature modulation observed here is consistent with the previous experiment[5] and the sign expected from the SSE[15].

Now we are in a position to investigate the spatial distribution of the SPE-induced temperature modulation. As already shown in Figs. 2 and 3, the SPE signals appear on the PM/YIG interface. Surprisingly, we found that, even under steady-state conditions, the temperature modulation due to the SPE remains confined near the PM/YIG interface without accompanying a temperature change on the areas away from the interface (see Fig. 4c and the temperature profiles measured with low lock-in frequency $f$ in Fig. 4e), a situation quite different from thermal diffusion expected from conventional heat sources. To highlight this anomalous behavior, we compare the temperature distribution induced by the SPE with that induced by Joule heating using the Pt/YIG sample. The temperature modulation induced by Joule heating can be measured by applying a d.c. offset to the rectangular a.c. current, although the Joule-heating contribution is eliminated in the SPE measurements because of the zero offset (Fig. 4a,b). We demonstrated that the Joule heating increases the temperature of the Pt layer irrespective of the $\mathbf{J}_c$ direction and the magnitude of the temperature modulation gradually decreases with the distance from the Pt layer due to thermal diffusion (Fig. 4d,f). This clear contrast between the SPE and Joule-heating signals suggests that the SPE induces non-trivial heat sources near the PM/YIG interfaces.

To clarify cross-sectional temperature distribution in the PM/YIG sample, we carry out numerical calculations by means of a two-dimensional finite element method. A model system used for the calculations consists of two Pt films on an YIG substrate, constructed on the basis of the Pt/YIG sample used for the above experiments (see Supplementary Note 2 for details). We calculated steady-state temperature distribution in the $x$-$z$ plane in the model system using a standard heat diffusion equation. The temperature distribution induced by Joule heating can be reproduced simply by setting a single heat source on each Pt. As shown in Fig. 5b,d, the single heat source exhibits isotropic temperature change, the magnitude of which gradually decreases with the distance from the Pt. The calculated in-plane temperature profile of the sample surface agrees with the observed Joule-heating-induced temperature distribution (compare Figs. 4f and 5f). However, it is obvious that such a standard heat source and its diffusion cannot reproduce the SPE-induced temperature modulation confined near the PM/YIG interfaces.

The spatial distribution of the SPE signals can be explained by assuming the presence of a dipolar heat source, a symmetric pair of positive and negative heat-source components, near the PM/YIG interface. Here, we calculated the temperature distribution in the Pt/YIG model system with setting dipolar heat sources on the Pt/YIG interfaces, where the polarity of the dipolar heat sources is reversed between the left and right interfaces according to the symmetry of the SPE (see the schematic illustration in Fig. 5c). As shown in Fig. 5a, the dipolar heat source exhibits anisotropic temperature change. We found that, when such dipolar heat sources are placed near the sample surface, the temperature change is confined in the vicinity of the source positions and the surface-side heat component, placed in Pt, determines the sign of the net temperature change on the interface (Fig. 5c and Supplementary Fig. 2a). This is because the surface-side component accumulates near the PM/YIG interface and generates a greater heating or cooling effect due to small heat dissipation at the sample surface, while the inner-side component, placed in YIG, is diluted due to large heat conduction toward the bulk of YIG. The obtained in-plane temperature profile of the sample surface well reproduces the SPE-induced temperature distribution (compare Figs. 4e and 5e and recall that, in the LIT measurements, the amplitude is defined in the range of $A > 0$ and the phase provides sign information). Importantly, this behavior appears only when the amounts of the positive and negative components of the dipolar heat sources are exactly the same as each other, where the net heat amount is macroscopically cancelled out; if the net heat amount is finite, the temperature change shows large thermal diffusion like the Joule-heating-induced temperature distribution (Supplementary Fig. 2b).

The clarification of the SPE-induced temperature distribution allows us to estimate the actual magnitude of the SPE signals. Because of the interfacially-confined temperature distribution, the magnitude of the SPE signals on the PM/YIG interface is much greater than that on the bare YIG surface even in the vicinity of the interface. In fact, the amplitude of the SPE signal per unit current density $j_c$ estimated from the data in Fig. 2d is $A/j_c = 4.7 \times 10^{-13}$ Km$^2$/A on the Pt/YIG interface, which is more than one order of magnitude greater than the value estimated from the data in ref. 5: $8.3 \times 10^{-15}$ Km$^2$/A on the YIG surface near the interface. This underestimation arises from the fact that only the temperature of the bare YIG surface was detected in the conventional experiments.

The SPE can be used as temperature controllers or modulators in spintronic devices. Owing to the interfacially-confined temperature distribution, the SPE enables pinpoint temperature manipulation, which cannot be realized by conventional methods with large thermal diffusion. In addition to the straightforward potential application, the SPE combined with the LIT technique is applicable to magnetic imaging and magnetometry, since SPE signals reflect interfacial magnetization distribution (see the center images of Fig. 2a,b, where the visualization of two-dimensional magnetic domain structure is demonstrated). The magnetic imaging based on the SPE is realized also owing to its tiny thermal diffusion, enabling the extraction of local magnetization information with high spatial resolution. We anticipate that the unique capability of the SPE will offer a new direction in spintronic applications and the thermal imaging of thermo-spin phenomena will lead to further development of spin-current physics.

**Methods**

**Sample fabrication.** The single-crystalline YIG with a thickness of 112 μm was grown on a single-crystalline Gd$_3$Ga$_5$O$_{12}$ (GGG) substrate with a thickness of 0.4 mm by a liquid phase epitaxy method. To improve the lattice matching between YIG and GGG, a part of Y in YIG is substituted by Bi; the exact composition of the YIG crystal is Bi$_{0.1}$Y$_{2.9}$Fe$_5$O$_{12}$. The YIG surface was mechanically polished with alumina slurry. The U-shaped Pt and W films with a thickness of 5 nm and a width of 0.2 mm were fabricated on the YIG by an rf magnetron sputtering method. The Al$_2$O$_3$ film with a thickness of 1 nm was prepared by an atomic layer deposition method using trimethylaluminum and ozone as a precursor and an oxidant, respectively. For the LIT measurements, the surface of the samples was coated with insulating black ink, which mainly consists of SiZrO$_4$, Cr$_2$O$_3$, and iron-oxide-based inorganic pigments.

**Acknowledgments** The authors thank J. Shiomi, T. Oyake, A. Miura, T. Kikkawa, D. Hirobe, G. E. W. Bauer and S. Maekawa for valuable discussions. This work was supported by PRESTO "Phase Interfaces for Highly Efficient Energy Utilization" from JST, Japan, Grant-in-Aid for Scientific Research (A) (No. 15H02012), Grant-in-Aid for Scientific Research on Innovative Area, "Nano Spin Conversion Science" (No. 26103005) from MEXT, Japan, NEC Corporation, the Noguchi Institute, and E-IMR, Tohoku University. S.D. is supported by JSPS through a research fellowship for young scientists (No. 16J02422). T.H. is supported by GP-Spin at Tohoku University.

**Author Contributions** K.U. and E.S. planned and supervised the study. S.D. and K.U. designed the experiments and prepared the samples. S.D. collected and analyzed the data. S.D. performed the numerical calculations with input from K.U and support from R.I. and T.H. The manuscript was prepared by K.U. All the authors discussed the results, developed the explanation of the experiments, and commented on the manuscript.

**Author Information** The authors declare no competing financial interests. Correspondence and requests for materials should be addressed to K.U. (kuchida@imr.tohoku.ac.jp).

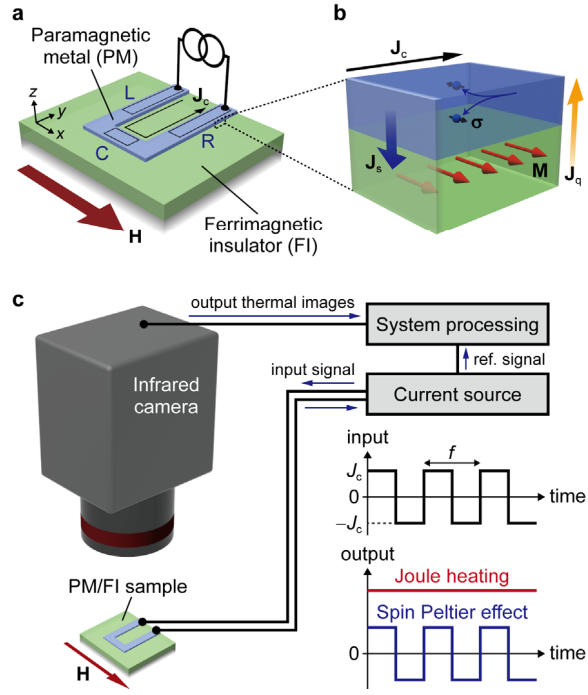

**Figure 1 | Device structure and experimental set-up. a,** A schematic illustration of the sample system used for measuring the spin Peltier effect (SPE). The sample comprises a U-shaped paramagnetic metal (PM; in experiments, Pt or W) film formed on a ferrimagnetic insulator (FI; in experiments, YIG). The squares on PM define the areas L, R, and C. **b,** The SPE induced by the spin Hall effect (SHE) near the PM/FI interface. **H**, **M**, $\mathbf{J}_c$, $\mathbf{J}_s$, and $\mathbf{J}_q$ denote the magnetic field vector (with the magnitude $H$), magnetization vector (with the magnitude $M$) of FI, charge current applied to PM, spatial direction of the spin current with the spin-polarization vector $\boldsymbol{\sigma}$ generated by the SHE in PM, and heat current generated by the SPE, respectively. Due to the symmetry of the SHE, the $\boldsymbol{\sigma}$ directions on L, R, and C are respectively along the $-x$, $+x$, and $-y$ ($+x$, $-x$, and $+y$) directions in Pt (W), the spin Hall angle of which is positive (negative). When **M** is along the $x$ direction, the SPE appears on L and R because of **M** ∥ $\boldsymbol{\sigma}$. **c,** Lock-in thermography (LIT) for the SPE measurements. When an a.c. charge current with rectangular wave modulation (with the amplitude $J_c$ and frequency $f$) is applied to PM, the SPE-induced temperature modulation ($\propto J_c$) oscillates with $f$, while the Joule-heating-induced temperature modulation ($\propto J_c^2$) is constant in time. The LIT system extracts the first harmonic response of observed thermal images, enabling the pure detection of the SPE free from the Joule-heating contribution.

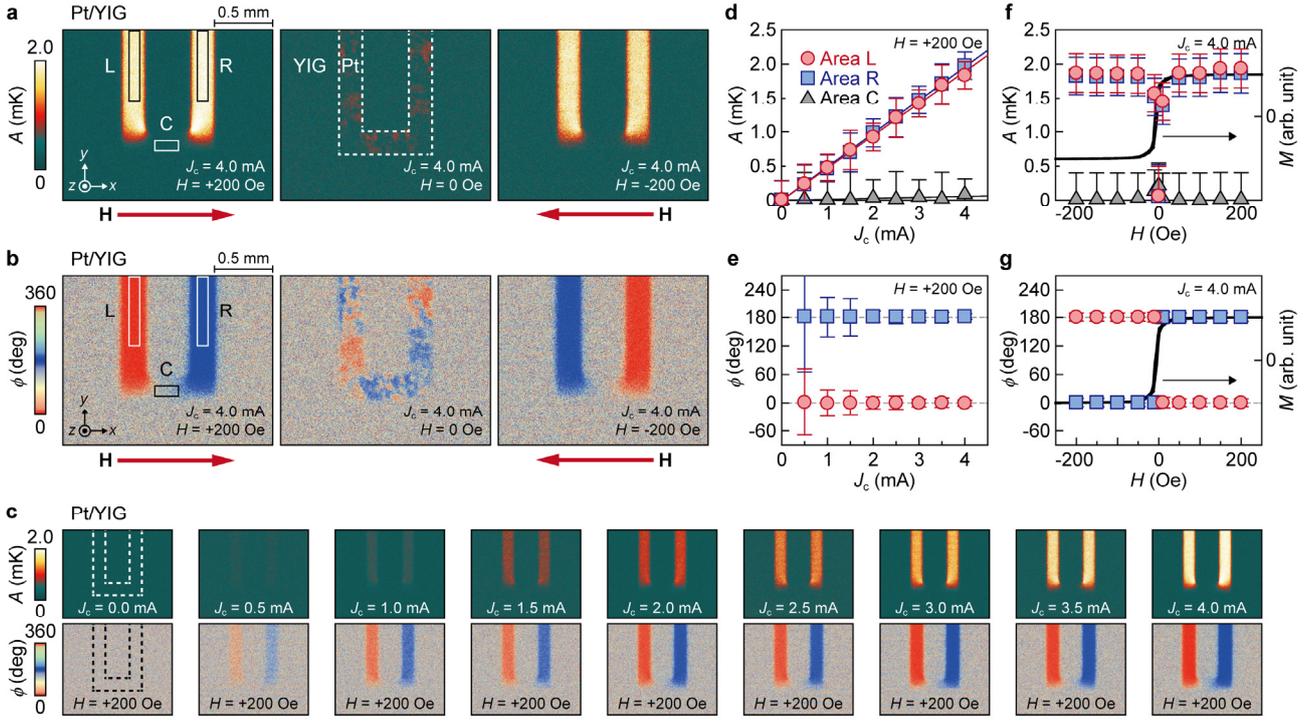

**Figure 2 | Observation of spin Peltier effect in Pt/YIG using lock-in thermography. a,b,** Lock-in amplitude $A$ (**a**) and phase $\phi$ (**b**) images for the Pt/YIG sample at $J_c = 4.0$ mA. The left, center, and right images were measured at $H = +200$ Oe, 0 Oe, and $-200$ Oe, respectively. **c,** $A$ and $\phi$ images for the Pt/YIG sample at $H = +200$ Oe for various values of $J_c$. **d,** $J_c$ dependence of $A$ on the areas L (red circles), R (blue squares), and C (gray triangles) of the Pt/YIG sample at $H = +200$ Oe. **e,** $J_c$ dependence of $\phi$ on L and R of the Pt/YIG sample at $H = +200$ Oe. **f,** $H$ dependence of $A$ on L, R, and C of the Pt/YIG sample at $J_c = 4.0$ mA and the $M$-$H$ curve (black line) of the YIG. **g,** $H$ dependence of $\phi$ on L and R of the Pt/YIG sample at $J_c = 4.0$ mA and the $M$-$H$ curve of the YIG. The data points in **d-g** are obtained by averaging the $A$ or $\phi$ values on L, R, and C, defined by the squares in the left images of **a** and **b**. The lock-in phase does not converge to a specific value when the signal amplitude is smaller than the sensitivity of the LIT; therefore, the $\phi$ data for C are not shown in **e** and **g**.

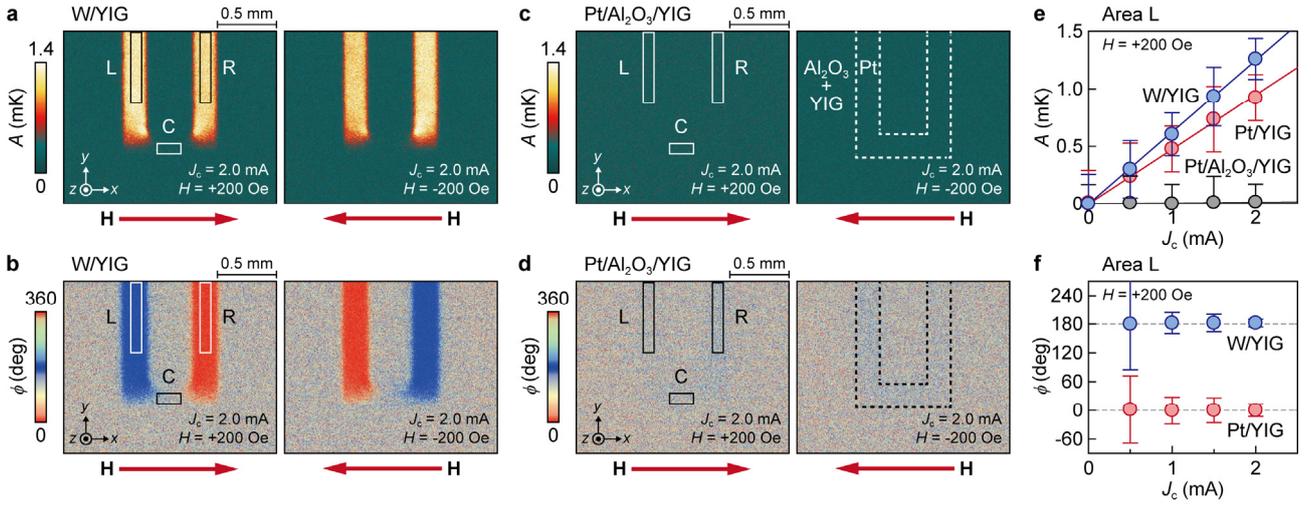

**Figure 3 | Thermal images in W/YIG and Pt/Al₂O₃/YIG. a,b,** $A$ (**a**) and $\phi$ (**b**) images for the W/YIG sample at $J_c = 2.0$ mA, measured at $H = +200$ Oe (left images) or $-200$ Oe (right images). **c,d,** $A$ (**c**) and $\phi$ (**d**) images for the Pt/Al₂O₃/YIG sample at $J_c = 2.0$ mA, measured at $H = +200$ Oe or $-200$ Oe. **e,** $J_c$ dependence of $A$ on the area L of the Pt/YIG (red circles), W/YIG (blue circles), and Pt/Al₂O₃/YIG (gray circles) samples at $H = +200$ Oe. **f,** $J_c$ dependence of $\phi$ on L of the Pt/YIG and W/YIG samples at $H = +200$ Oe.

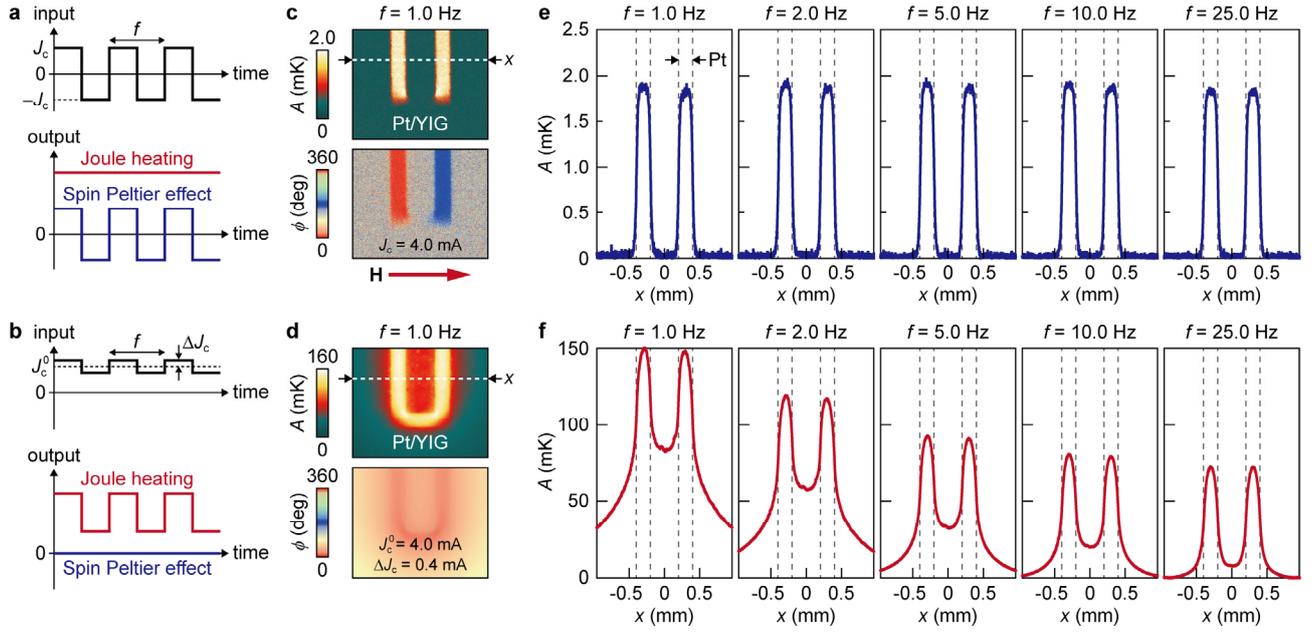

**Figure 4 | Temperature distributions induced by spin Peltier effect and Joule heating. a,b,** LIT conditions for the SPE (**a**) and Joule-heating (**b**) measurements. In the Joule-heating measurements, a d.c. offset of $J_c^0$ and an a.c. charge current with rectangular wave modulation (with the amplitude $\Delta J_c$ and frequency $f$) are applied to PM. In this condition, although both the SPE and Joule-heating signals appear in the first harmonic response of the thermal images, the observed LIT images are governed by the Joule-heating-induced temperature modulation because it is much greater than the SPE signal. **c,** $A$ and $\phi$ images for the Pt/YIG sample in the SPE condition (shown in **a**) at $J_c = 4.0$ mA, $H = +200$ Oe, and $f = 1.0$ Hz. **d,** $A$ and $\phi$ images for the Pt/YIG sample in the Joule-heating condition (shown in **b**) at $J_c^0 = 4.0$ mA, $\Delta J_c = 0.4$ mA, $H = 0$ Oe, and $f = 1.0$ Hz. **e,** One-dimensional $A$ profiles along the $x$ direction across the areas L and R of the Pt/YIG sample in the SPE condition at $J_c = 4.0$ mA and $H = +200$ Oe for various values of $f$. **f,** One-dimensional $A$ profiles along the $x$ direction across L and R of the Pt/YIG sample in the Joule-heating condition at $J_c^0 = 4.0$ mA, $\Delta J_c = 0.4$ mA, and $H = 0$ Oe for various values of $f$. The SPE-induced temperature distribution is independent of $f$, while the Joule-heating-induced temperature distribution broadens with decreasing $f$ due to thermal diffusion.

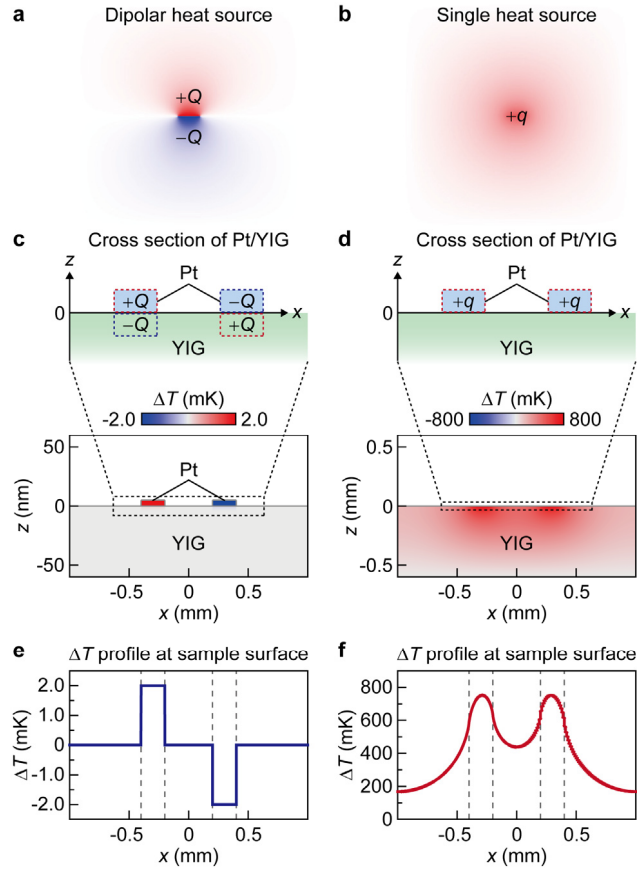

**Figure 5 | Numerical simulation of temperature distributions. a,b,** Calculated temperature distributions induced by a dipolar heat source comprising a symmetric pair of a positive component $+Q$ and a negative component $-Q$ (**a**) and a single heat source $+q$ (**b**). The finite element calculations were performed by using the COMSOL Multiphysics software. These distributions were obtained simply by setting the dipolar or single heat source at the center of a uniform YIG medium. **c,d,** Calculated temperature difference $\Delta T$ distributions induced by the dipolar heat sources on the Pt/YIG interfaces (**c**) and the single heat sources on the Pt (**d**) of the Pt/YIG model system. The design of the Pt/YIG system is detailed in Supplementary Note 2. The temperature of the bottom of the Pt/YIG system is fixed at 300 K and $\Delta T$ is defined as the difference from 300 K. Note that the calculation result in **c** is $10^4$ times magnified in the $z$ direction to emphasize the temperature distribution confined near the interface. **e,f,** One-dimensional $\Delta T$ profiles at the sample surface induced by the dipolar heat sources on the Pt/YIG interfaces (**e**) and by the single heat sources on the Pt (**f**) of the Pt/YIG system.

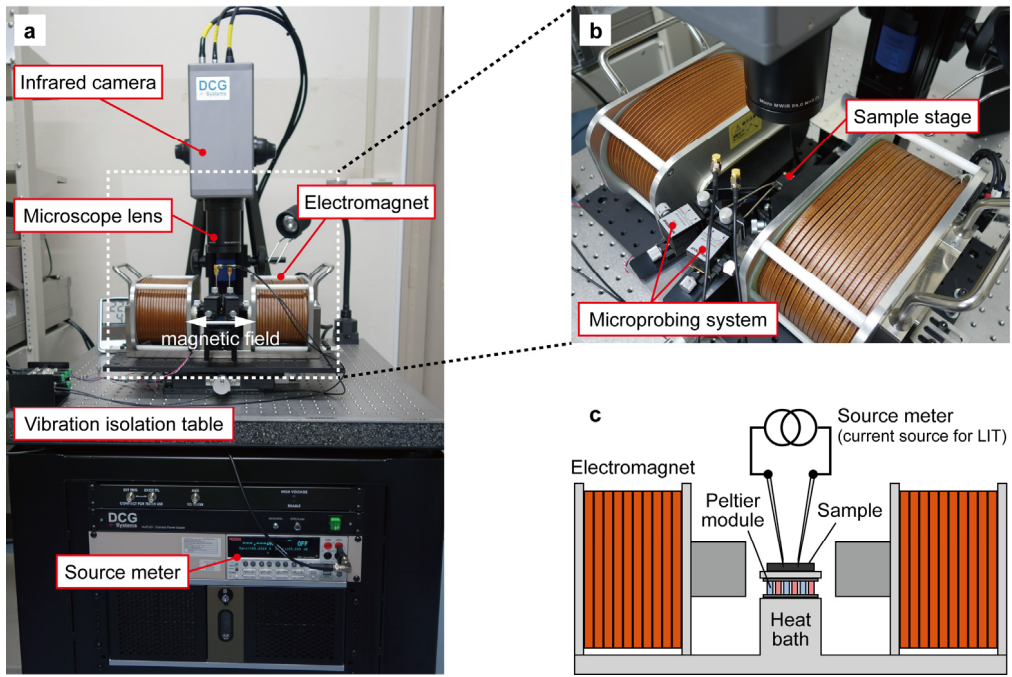

**Supplementary Figure 1 | Lock-in thermography system. a,** An overall picture of the lock-in thermography (LIT) system used in the present study. The LIT system (DCG Systems, ELITE) consists of an infrared camera with an InSb detector and a microscope lens, a system processing unit that performs real-time Fourier analysis of detected thermal images, and a source meter. The spectral range of the InSb detector is 3-5 μm. To measure thermal images under an in-plane magnetic field, an electromagnet is installed below the infrared camera. The camera and electromagnet are mounted on a vibration isolation table. **b,** A magnified view of the electromagnet and sample stage. During the LIT measurements, the Pt or W layer of the samples was connected to the output of the source meter via a microprobing system to apply a charge current. **c,** A schematic illustration of the electromagnet and sample stage. The temperature of the sample stage can be varied from 280 K to 320 K by using a thermoelectric Peltier module installed under the stage, where the Peltier module is mounted on a heat bath at room temperature. The temperature difference between the sample stage and the heat bath was measured with a differential thermocouple. This temperature control system was used for the calibration shown in Supplementary Note 1.

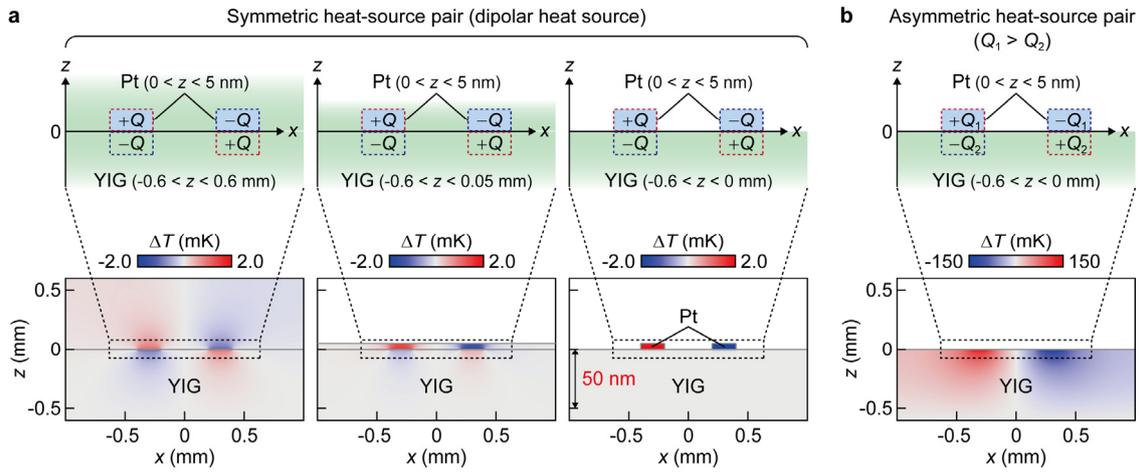

**Supplementary Figure 2 | Temperature distributions induced by symmetric and asymmetric pairs of positive and negative heat sources. a,** Calculated temperature difference $\Delta T$ distributions in the Pt/YIG model systems induced by the dipolar heat sources on the Pt/YIG interfaces for various YIG lengths along the $z$ direction. The dipolar heat source comprises a symmetric pair of a positive component $+Q$ and a negative component $-Q$. $\Delta T$ is defined as the difference from 300 K. The design of the Pt/YIG systems and the boundary conditions are detailed in Supplementary Note 2. We note that, in the Pt/YIG systems used for obtaining the left (center) result in **a**, the Pt is embedded in the YIG, where the YIG is in the range of $-0.6 < z < 0.6$ mm ($-0.6 < z < 0.05$ mm) and the Pt is in the range of $0 < z < 5$ nm. The other calculations were performed by using the Pt/YIG system in which the YIG is in the range of $-0.6 < z < 0$ mm and the Pt is put on the YIG surface. Note that the calculation result for the Pt/YIG system with the YIG of $-0.6 < z < 0$ mm is $10^4$ times magnified in the $z$ direction. **b,** Calculated $\Delta T$ distribution in the Pt/YIG model system induced by asymmetric pairs of positive and negative heat-source components on the Pt/YIG interfaces for the YIG of $-0.6 < z < 0$ mm. The asymmetric heat-source pair comprises the components $+Q_1$ ($-Q_1$) and $-Q_2$ ($+Q_2$) with $Q_1 = 1.01 \times Q_2$. Despite the small difference between $Q_1$ and $Q_2$, the calculated temperature distribution exhibits large thermal diffusion, which is clearly different from the temperature distribution induced by the dipolar heat sources and the experimental results in Fig. 4e. The temperature change is confined near the Pt/YIG interfaces only when $Q_1 = Q_2$.

**Supplementary Note 1 | Calibration method for lock-in thermography**

Infrared radiation intensity $I$ thermally emitted from the surface of materials depends on physical properties and surface conditions of the materials. Therefore, in the LIT experiments, the $I$ values detected by the infrared camera need to be converted into temperature $T$ information. This conversion is done by measuring the $T$ dependence of $I$. Since the LIT extracts thermal images oscillating with the same frequency as a periodic external perturbation applied to the sample, the $I$-to-$T$ conversion in the LIT is determined by the differential relation as $\Delta T_{1f}(\mathbf{r}) = dT/dI|_T \, \Delta I_{1f}(\mathbf{r})$, where $\Delta T_{1f}(\mathbf{r})$ and $\Delta I_{1f}(\mathbf{r})$ denote the first harmonic component of the temperature and infrared radiation intensity at the position $\mathbf{r}$, respectively.

In this study, we employed the following five-step calibration method.

(1) Measure the $T$ dependence of $I$ in the steady-state condition by using the infrared camera and the temperature control system shown in Supplementary Fig. 1c,

(2) Calculate the $dT/dI$ function from the obtained $I$-$T$ relation for each pixel,

(3) Perform the LIT measurements; measure the first harmonic response of the $I$ images, i.e. $\Delta I_{1f}$ images, with applying a periodic charge current to the sample,

(4) Determine $T$ values during the LIT measurements for each pixel by using the $I$-$T$ relation and steady-state $I$ images measured in parallel with the $\Delta I_{1f}$ images,

(5) Convert the $\Delta I_{1f}$ images into $\Delta T_{1f}$ images by applying the $dT/dI|_T$ value, obtained from the steps (2) and (4), to each pixel.

This calibration method is valid only when the infrared emissivity of the sample surface is very high. Therefore, the samples used in our experiments were coated with the black ink with the emissivity of $> 0.95$.

**Supplementary Note 2 | Procedures and conditions for numerical simulation**

The finite element calculations, shown in Fig. 5 and Supplementary Fig. 2, were performed by means of the COMSOL Multiphysics software. By using the following model systems and boundary conditions, we calculated the steady-state cross-sectional temperature distribution based on a standard heat diffusion equation.

The model system used for the calculations in Fig. 5a,b is a simple YIG medium with a 2 mm × 2 mm square shape. The temperature of the four sides of the YIG square is fixed at 300 K as a boundary condition. In Fig. 5a (5b), we set a dipolar (single) heat source at the center of the YIG square, where the size of the positive and negative components of the dipolar heat source (the size of the single heat source) is 0.2 mm × 5 nm. The thermal conductivity, density, and specific heat of YIG are assumed to be the values shown in Supplementary Table 1.

The Pt/YIG model system used for the calculations in Fig. 5c,d and Supplementary Fig. 2b consists of two Pt rectangles put on an YIG rectangle. The lengths of the YIG rectangle (each Pt rectangle) along the $x$ and $z$ directions are 2 mm (0.2 mm) and 0.6 mm (5 nm), respectively. Here, the bottom of the Pt rectangles is fixed at $z$ = 0. The distance between the centers of the Pt rectangles along the $x$ direction is 0.6 mm. To reproduce the experimental situations, we set the following boundary conditions. The top and side surfaces of the Pt/YIG system are connected to air and the temperature of the bottom surface of the YIG is fixed at 300 K. The thermal conductivity, density, and specific heat of each component are assumed to be the values shown in Supplementary Table 1. The interfacial thermal resistance at the Pt/YIG interfaces and the heat-transfer coefficient from the Pt/YIG system to air are set to be 2.79 × 10$^8$ W/m$^2$K (ref. S1) and 10 W/m$^2$K (ref. S2), respectively. In the calculations in Fig. 5c, we set dipolar heat sources on the Pt/YIG interfaces, where one of the components of the dipolar heat sources is placed in the Pt rectangles and the other is in the YIG rectangle. In the calculations in Fig. 5d, we set single heat sources with the same sign in the Pt rectangles. The size of the positive and negative components of the dipolar heat sources and the single heat sources is 0.2 mm × 5 nm. Our numerical calculations confirmed that the macroscopic temperature distributions induced by the dipolar heat sources do not change qualitatively even when the material parameters and the position and size of the heat sources are varied within reasonable ranges. We also checked that the surface temperature profiles of the Pt/YIG system is maintained even when a black-ink coating with a thickness of several tens of micrometers is attached to the Pt/YIG surface.

The numerical calculations in Supplementary Fig. 2a were performed under the same conditions as those used for Fig. 5c except for the YIG length along the $z$ direction. As described in the caption of Supplementary Fig. 2, in the Pt/YIG systems with the YIG of −0.6 < $z$ < 0.6 mm and −0.6 < $z$ < 0.05 mm, the Pt rectangles are embedded in the YIG. The YIG-length dependence of the temperature distribution clearly shows that, when the dipolar heat sources are placed near the sample surface, the temperature change is confined in the vicinity of the source positions.

**Supplementary Table 1 | Material parameters for numerical simulation.**

|  | Pt | YIG |
|---|---|---|
| Thermal conductivity (W/mK) | 72 (ref. S3) | 7.4 (ref. S3) |
| Density (kg/m$^3$) | 21450 (ref. S1) | 5170 (ref. S1) |
| Specific heat (J/kgK) | 132.56 (ref. S1) | 570 (ref. S1) |

**Supplementary References**